\UseRawInputEncoding
\documentclass[9pt, 
superscriptaddress, groupedaddress, preprintnumbers, nofootinbib, amsmath, amssymb, aps, prd, twocolumn]{revtex4-2}

\usepackage{orcidlink}
\usepackage{amsfonts}
\usepackage[nolist,nohyperlinks]{acronym}
\usepackage{bookmark}
\usepackage{xcolor}
\usepackage{graphicx}
\usepackage{dcolumn}
\usepackage[version=4]{mhchem}
\usepackage{bm}
\usepackage{braket}
\usepackage{siunitx}
\usepackage{array}

\begin{document}

\preprint{APS/123-QED}

\title{\textbf{Dark matter self-annihilation as a hidden energy source in white dwarfs}}


\author{Alvin Cheuk-Nam Chu\orcidlink{0009-0001-7807-5809}}
\email{alvin.cnchu@link.cuhk.edu.hk}
\affiliation{Department of Physics and Institute of Theoretical Physics, The Chinese University of Hong Kong, Shatin, N.T., Hong Kong S.A.R., China}

\author{Cheuk-Man Yiu\orcidlink{0009-0008-7518-5309}}
\affiliation{Department of Physics and Institute of Theoretical Physics, The Chinese University of Hong Kong, Shatin, N.T., Hong Kong S.A.R., China}

\author{Ching-Hui Lam\orcidlink{0009-0000-3994-0284}}
\affiliation{Department of Physics and Institute of Theoretical Physics, The Chinese University of Hong Kong, Shatin, N.T., Hong Kong S.A.R., China}

\author{Peter Siu-Hei Cheung\orcidlink{0000-0002-0814-3378}}
\affiliation{Zentrum f\"{u}r Astronomie der Universit\"{a}t Heidelberg, Astronomisches Rechen-Institut, M\"{o}nchhofstr. 12-14, 69120 Heidelberg, Germany}
\affiliation{Heidelberger Institut f\"{u}r Theoretische Studien, Schloss-Wolfsbrunnenweg 35, 69118 Heidelberg, Germany}
\affiliation{Department of Physics and Institute of Theoretical Physics, The Chinese University of Hong Kong, Shatin, N.T., Hong Kong S.A.R., China}

\author{Fong-Ching Ho\orcidlink{0009-0001-6253-6287}}
\affiliation{Department of Physics and Institute of Theoretical Physics, The Chinese University of Hong Kong, Shatin, N.T., Hong Kong S.A.R., China}

\author{Ming-chung Chu\orcidlink{0000-0002-1971-0403}}
\affiliation{Department of Physics and Institute of Theoretical Physics, The Chinese University of Hong Kong, Shatin, N.T., Hong Kong S.A.R., China}

\date{\today}


\begin{abstract}
\acp{WD} are compact remnants of stars, supported by electron degeneracy pressure, and their structures are generally considered to be well understood. Significant discrepancies exist between the theoretical and observed white dwarf \ac{MR} relations, especially for hot Q-branch white dwarfs with unexplained heating sources. In this study, we explore the potential of \acp{WD} as a portal to \ac{DM} physics by investigating the effect of \ac{SADM} admixed in \acp{WD}, solving the set of two-fluid hydrostatic equilibrium and heat diffusion equations self-consistently. We obtain empirical formulae for the \ac{WD} luminosity $L$ and effective temperature $T_\text{eff}$ as functions of the \ac{DM} mass fraction $f_d$, annihilation cross section $\braket{\sigma v}$, and particle mass $m_\chi$, as well as the \ac{WD}'s total mass $M_t$. We also compare the \ac{WD} \ac{MR} relations for various $L$, $f_d$, and both Fermionic and Bosonic \ac{DM} particle statistics, with the observational data. We show that the majority of observed deviations in the \ac{WD} \ac{MR} relation can be accounted for by having \ac{SADM} admixed with $f_d\leq10^{-5}$ and $\braket{\sigma v}\ll3\times 10^{-26}~\si{cm}^3~\si{s}^{-1}$, the thermal relic cross section. Therefore, our results suggest that \acp{WD} can be sensitive \ac{DM} detectors, and \ac{DM} self-annihilation could provide additional heating in \acp{WD}, accounting for their observed extra luminosities.

\end{abstract}


\maketitle

\begin{acronym}
\acro{MR}{mass-radius}
\acro{NM}{normal matter}
\acro{DM}{dark matter}
\acro{FDM}{fermionic \ac{DM}}
\acro{BDM}{bosonic \ac{DM}}
\acro{SADM}{self-annihilating \ac{DM}}
\acro{WD}{White dwarf}
\acro{EoS}{equation of state}
\acro{MS}{main-sequence}
\end{acronym}

\acresetall


\section{Introduction}
\begin{figure}
\includegraphics[width=\linewidth]{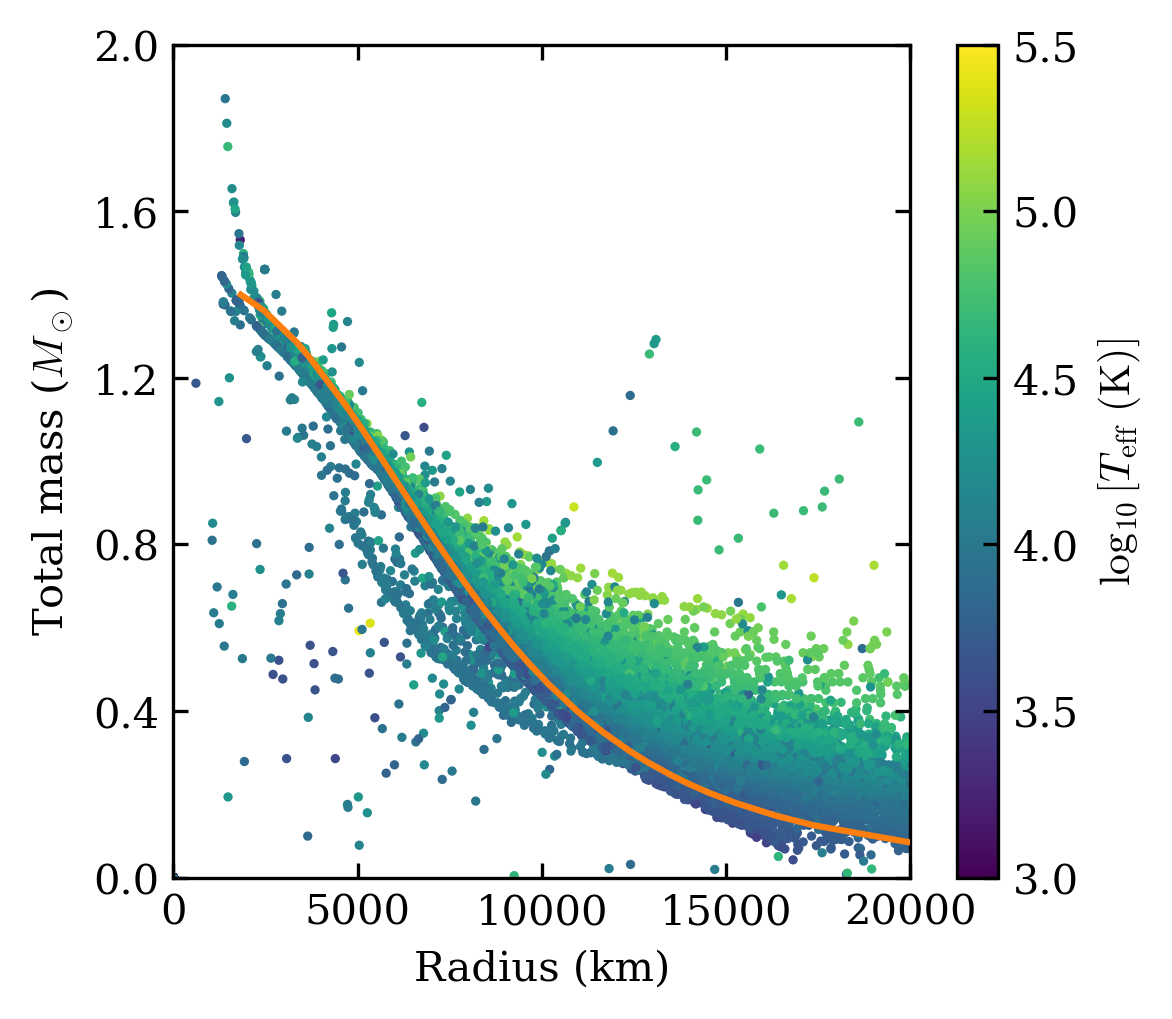}
\caption{\label{fig:MRT}\ac{WD} \ac{MR} relations. Observational data from the Montreal \ac{WD} Database \cite{dufour2016montrealwhitedwarfdatabase} are shown as dots, with the colors representing their $T_\text{eff}$. The theoretical \ac{MR} relation for pure non-rotating cold carbon-oxygen \acp{WD} is plotted in orange.}
\end{figure}

\acp{WD} are supported primarily by electron degeneracy pressure and cool over time without internal nuclear fusion \cite{doi:https://doi.org/10.1002/9783527617661.ch3}, rendering their standard evolution conceptually straightforward. However, many observed \acp{WD} \cite{dufour2016montrealwhitedwarfdatabase, GaiaDR2} deviate from the theoretical \ac{MR} relation as their effective surface temperatures ($T_\text{eff}$) increases (see FIG.~\ref{fig:MRT}). In particular, some \acp{WD} are hot but old (effective temperature $>10 000$ K at cooling age $>5$ Gyr), but they are expected to cool to thousands of Kelvin millions of years after their formation. Previous studies of hot \acp{WD} have shown that temperature effects modify the theoretical \ac{MR} curve to agree with observations \cite{Silvia_2021, Crumpler_2024}. Moreover, Tremblay et al. discovered the so-called Q-branch \acp{WD} characterized by a pile-up in the cooling sequence \cite{Tremblay2019, GaiaDR2}. Cheng et al. argued that the standard crystallization model cannot fully explain this branch, suggesting additional heating from an unknown energy source \cite{Cheng_2019}. Extra energy sources inside \acp{WD} have been suggested, such as hydrogen accretion from the binary companion or interstellar matter \cite{10.1093/mnras/112.6.583}, gravitational settling of \ce{^{22}Ne} \cite{Bildsten_2001, Blouin_2021}, and proton decay catalyzed by magnetic monopoles \cite{liu2024}, etc., which would heat and enlarge the \acp{WD}. 

Many observations, such as galactic rotation curves \cite{Rubin1980RotationalPO, 10.1046/j.1365-8711.2000.03075.x} and cosmic microwave background anisotropies \cite{damico2009darkmatterastrophysics}, support the existence of \ac{DM}, which is believed to contribute approximately one-fourth of the energy density of the universe today \cite{peter2012darkmatterbriefreview}. It is therefore natural to expect that \ac{DM} could be admixed in stars, since their formation and throughout their entire evolution \cite{Iocco_2008}. \acp{WD} could also contain \ac{DM} \cite{Chan_2022, moskalenko2007darkmatterburnerspreliminary}, modifying their \ac{MR} relation and making them natural laboratories for \ac{DM} search.

\ac{DM} interacts with \ac{NM}, the Standard Model particles, via gravity. However, there is no general consensus on \ac{DM} properties such as mass, spin, and interactions. Being electrically neutral, \ac{DM} could be Majorana particles that self-annihilate \cite{JUNGMAN1996195, BERTONE2005279}. Such \ac{SADM} has been actively studied for decades. In \cite{refId0}, it was found that \ac{SADM} admixed into low-mass red giant branch stars would generate energy and raise their luminosity.

Ongoing research examines \ac{DM} admixed compact objects. It was shown that \acp{WD}' observables, e.g., mass, radius, etc., could be altered by the presence of \ac{FDM} \cite{PhysRevD.87.123506} or \ac{BDM} \cite{PhysRevD.109.095032,MOSQUERA2010119} admixture. Chan et al. demonstrated that rotating DM admixed \acp{WD} could account for some peculiar \acp{WD} \cite{Chan_2022}.  Furthermore, \ac{DM} admixture could also induce changes in other \ac{WD} observables, such as the luminosity function \cite{Bertolami_2014} and the surface temperature \cite{PhysRevD.77.043515}. It was even shown in \cite{PhysRevD.87.055012} that self-annihilating \ac{BDM} could prevent old neutron stars from collapsing. 

To tackle the extra heating,  Horowitz explored the possibility of \ac{DM} annihilation as the energy source with a simple model \cite{PhysRevD.102.083031}. We are thus motivated to study \ac{SADM} in detail as a candidate extra energy source in \acp{WD} that would alter their observables.

Constraints on the \ac{DM} annihilation cross section $\braket{\sigma v}<3\times10^{-26}$ cm$^3$ s$^{-1}$ \cite{PhysRevD.86.023506} have been derived from the cooling time and tidal deformability of neutron stars \cite{Mariani_2023,PhysRevD.110.023024} and dwarf galaxy structure \cite{PhysRevD.106.075007}. Moreover, constraints on \ac{DM}-nucleon interaction cross section were obtained by studying \ac{SADM}-admixed \acp{WD} \cite{Acevedo_2024}. Here, given known constraints, we explore the possibility of \ac{SADM} producing observable effects on \ac{SADM}-admixed \acp{WD}.

In Section~\ref{sec:Methodology}, we discuss how we formulate and obtain the hydrostatic solution for \ac{SADM}-admixed \acp{WD}. Then in Section~\ref{sec:Results}, we show the relationship between temperature and the two-fluid density profile. We report the empirical formulae obtained for the \ac{WD} luminosity and effective temperature from \ac{SADM} with \ac{DM} parameters. We also show the possibility for explaining the observed discrepancies in the \ac{WD} \ac{MR} relation with \ac{SADM}-admixed \acp{WD}. We conclude our findings in Section~\ref{sec:Conclusion}. In Appendix~\ref{app:capture rate}, we estimate the \ac{DM} capture rate of a \ac{WD}.


\section{Methodology} \label{sec:Methodology}

\subsection{Spherically symmetric two-fluid hot \acp{WD} in equilibrium}
We first consider a spherically symmetric two-fluid \ac{WD}. The \ac{WD} is described by a set of two coupled general relativistic hydrostatic equations, the Tolman-Oppenheimer-Volkoff equations \cite{Tolman1939, Oppenheimer1939}
\begin{align}
    \frac{dP_n}{dr}&=-\frac{G}{r^2}\frac{(\rho_n+P_n)(M_t+4\pi r^3 P_{t})}{(1-\frac{2GM_t}{r})},\\
     \frac{dP_d}{dr}&=-\frac{G}{r^2}\frac{(\rho_d+P_d)(M_t+4\pi r^3 P_{t})}{(1-\frac{2GM_t}{r})}.
\end{align}

Here, the subscripts $n$, $d$, and $t$ denote \ac{NM}, \ac{DM}, and the sum of them, respectively, and $\rho$ is the rest-mass density, $M$ the enclosed mass, $P$ the pressure, and $r$ the radial position. We set $c=\hbar=k_B=1$ while keeping the universal gravitational constant $G$ explicit.

We adopt the Helmholtz \ac{EoS} for the \ac{NM} pressure $P_n(\rho_n,T)$ \cite{1999ApJS..125..277T, 2000ApJS..126..501T}, where $T$ is the \ac{NM} temperature. We pick \ce{^{12}C} as our \acp{WD}' main component to compute $P_n$, which allows the modeling of a hot WD system. We ignore Coulomb corrections to ensure thermodynamic consistency.

For the system to remain in equilibrium, the energy dissipation, i.e., the outward energy flux $F$, balances the total luminosity $L$ enclosed. Thus, we have the diffusion equation
\begin{equation} \label{eq:diffusion}
    F=\frac{L}{4\pi r^2}=-\frac{16\sigma T^3}{3\kappa\rho_n}\frac{dT}{dr},
\end{equation}
with the boundary condition at the \ac{NM} surface $r=R_n$
\begin{equation}
    L(R_n)=4\pi R_n^2\sigma T(R_n)^4,
\end{equation}
where $\sigma$ is the Stefan-Boltzmann constant and $\kappa$ is the effective opacity. This allows us to determine the temperature profile $T(r)$ of our \acp{WD}.

The effective opacity $\kappa$ in Eq.~\eqref{eq:diffusion} determines the efficiency of energy absorption, given by
\begin{equation}
    \kappa=\frac{\kappa_R}{1+\lambda_C/\lambda_R},
\end{equation}
where $\kappa_R$ is the radiational opacity, and $\lambda_R$ and $\lambda_C$ are the radiational and thermal conductivities, respectively. We adopt the expressions for both degenerate and non-degenerate cases from Table I in \cite{Kothari1932ApplicationsOD}. For the degenerate case, we have
\begin{align} 
    \kappa_R &= \frac{28}{15\pi\sqrt{3}}\frac{e^6Z^2}{m_HA'}\frac{1}{T^2},\\
    \lambda_R &= \frac{16\sigma}{3}\frac{T^3}{\kappa_R\rho_n},\\
    \lambda_C &= \pi^4\frac{Y_e}{e^4 m_e^2 m_b Z}\frac{T\rho_n }{ I_1},\\
    I_1 &= 2\pi \ln\left(1+\left(5.55\times10^{-2}\left(\frac{\rho_n}{1~\si{g}~\si{cm}^{-3}}\right)^{\frac{1}{3}}\right)^2\right).
\end{align}

For the non-degenerate case, we have
\begin{align}
    \kappa_R &= \frac{32\pi^4(1.0823)}{315\sqrt{3}(1.0128)}\frac{e^6}{(2\pi m_e)^{3/2}m_H}\frac{Z^2Y_e}{A'm_b}\frac{\rho_n}{T^{7/2}},\\
    \lambda_R &= \frac{16\sigma}{3}\frac{T^3}{\kappa_R\rho_n},\\
    \lambda_C &= \frac{2^{13/2}}{\pi^{1/2}}\frac{T^{5/2}}{e^4 m_e^{1/2} ZI_1},\\
    I_1 &= 2\pi \ln\left(1+\left(8.45\times10^{-7}\left(\frac{T}{1~\si{K}}\right)\left(\frac{1~\si{g}~\si{cm}^{-3}}{\rho_n}\right)^{\frac{1}{3}}\right)^2\right).
\end{align}

Here, $e$ is the elementary charge, and $Y_e$ the electron fraction. $Z$ and $A'$ are the atomic and mass numbers, respectively, which depend on the composition of the \ac{WD}, and we take \ce{^{12}C} for the computation of opacity; $m_H, m_e$, and $m_b$ are the rest masses of a hydrogen atom, electron, and baryon (defined as $1/12$ the mass of a \ce{^{12}C} atom), respectively.


\subsection{Fermionic \ac{DM} \ac{EoS}}

We assume the \ac{FDM} to form a degenerate fermi gas at the centre of the \ac{WD}, supported by its own degeneracy pressure, described by the zero-temperature ideal Fermi gas \ac{EoS} \cite{doi:https://doi.org/10.1002/9783527617661.ch3,2007coaw.book.....C, PhysRevD.87.123506}:
\begin{align}
    P_d&=\frac{B}{8}\left(x(2x^2-3)\sqrt{x^2+1}+3\sinh^{-1}x\right),\\ 
    x&=\left(\frac{\rho_d}{B}\right)^{\frac{1}{3}},~B=\frac{m_\chi^4}{3\pi^2}.
\end{align}
Here, $m_\chi$ is the \ac{DM} particle mass.


\subsection{Bosonic \ac{DM} \ac{EoS}} 
We assume the self-interacting \ac{BDM} condenses into a Bose-Einstein condensate \cite{Böhmer_2007, PhysRevD.84.043531}, characterized by a single wave function $\psi$. The wave function is described by the ground state solution of the Gross-Pitaevskii equation \cite{Gross1961,1571698599883889664} as 
\begin{equation}
    \left(-\frac{1}{2m_\chi}\nabla^{2}+m_\chi\Phi+\frac{4\pi a_s}{m_\chi^2}\rho_d\right)\psi=i\frac{\partial\psi}{\partial t},
    \label{eq:GP}
\end{equation}
with the Newtonian gravitational potential $\Phi$, the scattering length $a_s$, where $a_s$ $>$ ($<$) 0 indicates repulsive (attractive) interaction between \ac{BDM} particles. Eq.~\eqref{eq:GP} can be transformed into a hydrodynamical equation by applying the Madelung transformation \cite{Madelung}, with the pressure of the \ac{BDM} Madelung fluid given by 
\begin{equation}
    P_d=\frac{2\pi a_s}{m_\chi^3}\rho_d^2,
\end{equation}
which is a polytropic \ac{EoS} \cite{PhysRevD.84.043531}. We focus on massive \ac{BDM} with $m_\chi\geq0.01~\si{GeV}$, for which, the quantum potential term $\frac{1}{2m_\chi^2}\nabla\left(\frac{\nabla^2\sqrt{\rho_d}}{\sqrt{\rho_d}}\right)$ is negligible \cite{PhysRevD.84.043531, RevModPhys.71.463}.


\subsection{Self-annihilating dark matter}
We consider the \ac{SADM} particle $\chi$ to be self-annihilating into two Standard Model photons $\gamma$ ($\chi\chi\rightarrow \gamma\gamma$) per interaction \cite{BERTONE2005279, BERGSTROM199727}, each carrying energy $m_\chi$. The power per unit volume $\epsilon_{\chi\chi}$ produced by the self-annihilation is given by 
\begin{equation}
    \epsilon_{\chi\chi}=\frac{1}{2}\braket{\sigma v}\left(\frac{\rho_d}{m_\chi}\right)^2(2 m_\chi),
\end{equation}
where $\braket{\sigma v}$ is the velocity-weighted cross section of \ac{DM}-\ac{DM} annihilation and the interaction rate is proportional to the \ac{DM} particle number density squared. $2m_\chi$ is the energy produced per interaction. The prefactor $1/2$ accounts for identical particles.

Integrating $\epsilon_{\chi\chi}$ over the volume of a region gives the enclosed luminosity
\begin{equation}
    L=\int_V dV~\epsilon_{\chi\chi}. 
\end{equation}
With \ac{DM} self-annihilation, the produced photons scatter and transfer energy to the \ac{NM}, and the energy then diffuses into the star. This gives rise to the \ac{WD} temperature described by Eq.~\eqref{eq:diffusion}.


\subsection{Temperature constraints}
In the high density and temperature region, nuclear reactions and electron capture could be triggered, leading to an unstable core.

\begin{figure}
\includegraphics[width=\linewidth]{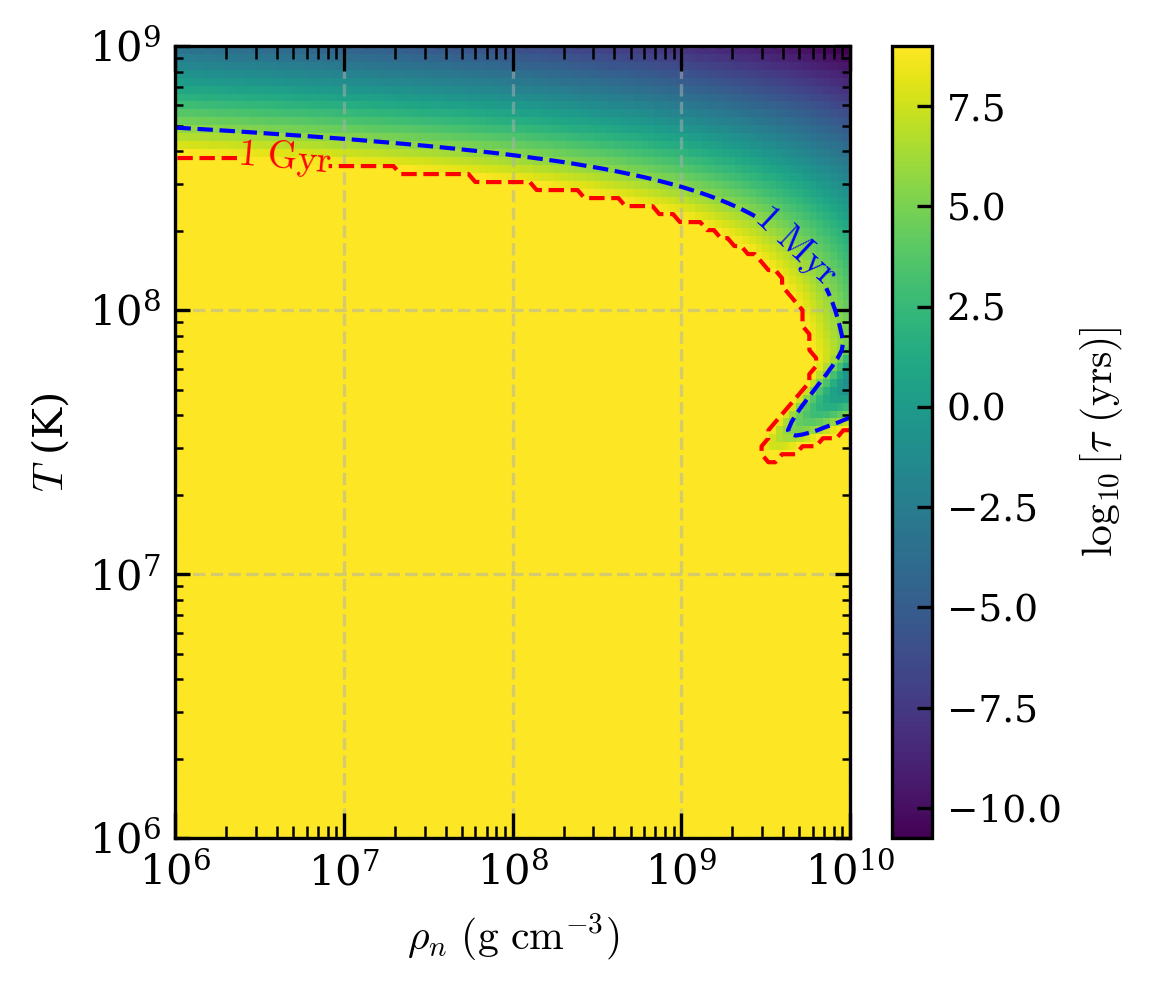}
\caption{\label{fig:C-12 depletion timescale} The time to 90\% abundance $\tau$ at different initial density and temperature. The red (blue) dashed line indicates where $\tau$ equals to 1 billion (million) years.}
\end{figure}

To determine the stability of a $^{12}\mathrm{C}$ core, we construct a nuclear reaction network using \texttt{pynucastro} \cite{pynucastro2}, similar to the one in \cite{2022ApJ...936....6Z, Zingale_2024}, with the addition of the 
\ce{^{20}Ne (e^-, $\nu_{\text{e}}$) ^{20}F (e^-, $\nu_{\text{e}}$) ^{20}O} and \ce{^{24}Mg (e^-, $\nu_{\text{e}}$) ^{24}Na (e^-, $\nu_{\text{e}}$) ^{24}Ne} sequences. We obtain the nuclear reaction rates from the ReacLib database \cite{Cyburt_2010} and the weak interaction rates from \cite{Suzuki_2016}. The effect of screening \cite{PhysRevD.76.025028} is also included in this analysis. We sample the temperature and density on a $100 \times100$ grid, logarithmically spaced in $\rho_n\in[10^6,10^{10}]$ \si{g.cm^{-3}} and $T \in [10^6, 10^9]$ K. We use the BDF integrator from \texttt{SciPy} \cite{2020SciPy-NMeth} to evolve the nuclear network. We solve for the depletion timescale $\tau$ at which the \ce{^{12}C} mass fraction $X_{\ce{^{12}C}}$ reaches 0.9 from 1 initially, and the results are shown in FIG.~\ref{fig:C-12 depletion timescale}.

We limit the NM central density-temperature combination of our solutions so that a pure \ce{^{12}C} core maintains at least a 90\% abundance for a billion years.

\subsection{Stability}
We perform a linear hydrodynamic stability test on the hydrostatic solutions following the treatment in \cite{Ledoux1950TheVS}, by introducing a displacement perturbation $\delta r$ to the equilibrium fluids. The condition of stability is given by
\begin{equation}\label{eq:stability condition}
\int_{0}^{M_n}(\Gamma_T - 1)\frac{\delta\rho_n}{\rho_n}\delta\left(\frac{\epsilon_{\chi\chi}}{\rho_n}-\nabla\cdot F\right)dm<0,
\end{equation}
where $\delta(\cdot)$ denotes the co-moving variations of dynamic quantities resulting from the perturbation, $\Gamma_T=\frac{1}{\rho_n}(\frac{\partial P_n}{\partial E})_{\rho_n}$ is the adiabatic index, and the integral is over all the mass shells in the \ac{WD}. The radial perturbation $\xi\equiv\frac{\delta r}{r}$ is small and we follow \cite{1949AnAp...12...39S} to approximate it as a constant throughout the star. We then apply the linear perturbation relations to Eq.~\eqref{eq:stability condition} following \cite{Ledoux1950TheVS}
\begin{align}
    \frac{\delta\rho_n}{\rho_n}&=-3\xi,\\
    \frac{\delta T}{T}&=(\Gamma_T-1)\frac{\delta\rho_n}{\rho_n}.
\end{align}
We rule out solutions violating Eq.~\eqref{eq:stability condition} and include only stable solutions in our analysis.


\section{Results} \label{sec:Results}

\subsection{Density and temperature profiles of a \ac{SADM}-admixed \ac{WD}}

\begin{figure}
    \centering
    \includegraphics[width=\linewidth]{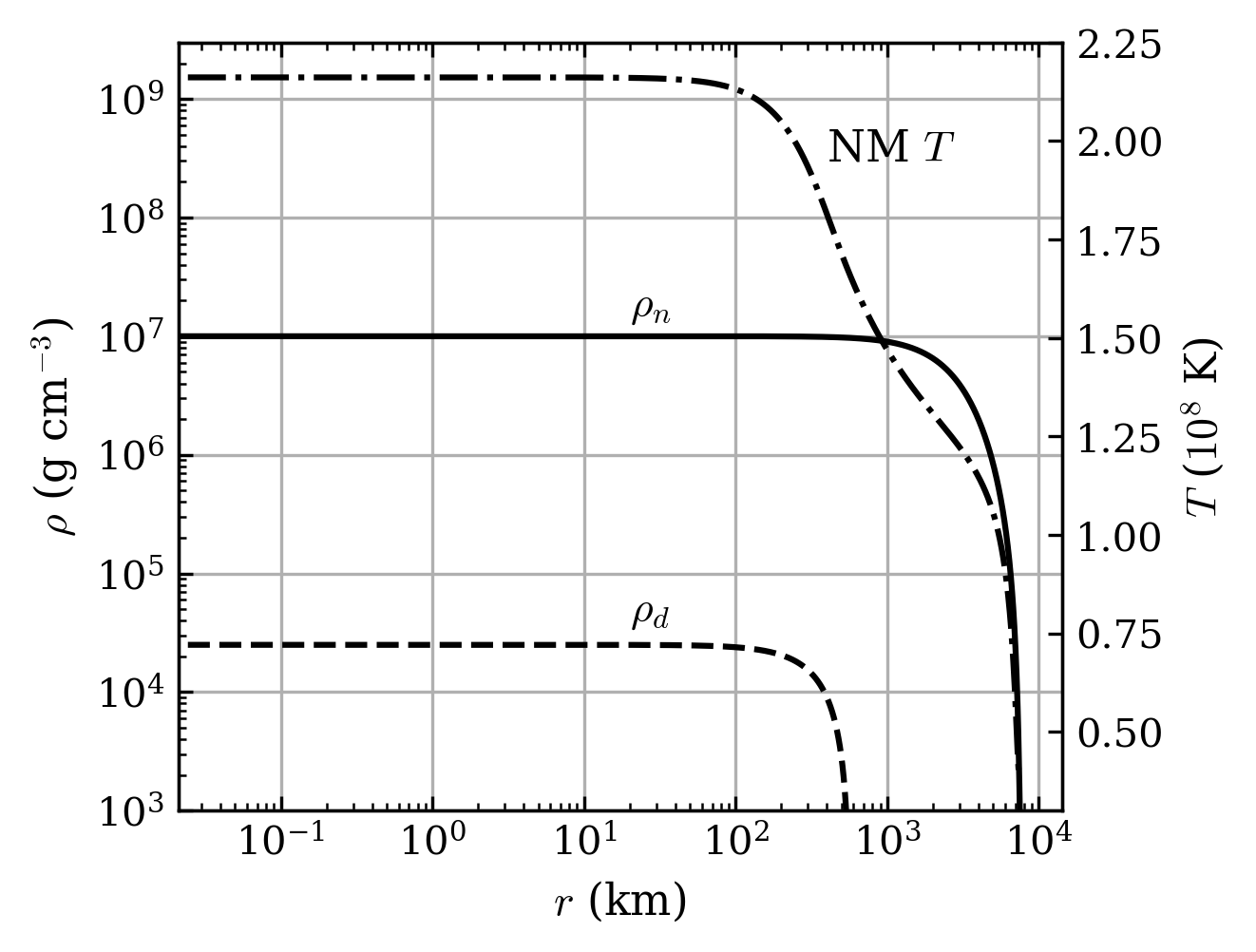}
    \caption{\label{fig:Trho profile}\ac{NM} (solid line) and \ac{FDM} (dashed line) density profiles together with the \ac{NM} temperature profile (dot-dashed line) against $r$ in a typical \ac{SADM}-admixed \ac{WD} in log-scale for both axes, with \ac{FDM} parameters $\braket{\sigma v}=10^{-44}$ cm$^3$ s$^{-1}$, $m_\chi=0.1$ GeV. The central densities for \ac{NM} and \ac{DM} are $\rho_{c,n}=10^7$ g cm$^{-3}$ and $\rho_{c,d}=2.5\times10^4$ g cm$^{-3}$, respectively.}
\end{figure}

The \ac{NM} is heated by \ac{DM} annihilation in a \ac{SADM}-admixed \ac{WD}, and we plot its typical temperature and density profiles in FIG.~\ref{fig:Trho profile}. The \ac{NM} is isothermal inside the \ac{DM} core due to the high thermal conductivity of the \ac{WD} core.

As the \ac{DM} density decreases, leading to less radiation energy, the equilibrium temperature profile falls off. Outside the \ac{DM} fluid, with no more energy sources, $T$ decreases gradually with the \ac{NM} density.

We only consider $m_\chi\in[0.01,~10]~\si{GeV}$ so that the \ac{DM} cores sit inside the \acp{WD}. For different ratios of \ac{NM}-\ac{DM} radii, the $T$ profiles are similar in shape, and the temperature is within the same order of magnitude throughout the stars.


\subsection{Empirical fitting of \ac{SADM}-admixed \acp{WD}' luminosity}

We obtain an empirical fitting formula of the \ac{SADM}-admixed \ac{WD} luminosity $L$ as
\begin{equation}
\left(\frac{L}{L_\odot}\right)\left(\frac{M_\odot}{M_t}\right)^{3.5}=\left\{
    \begin{aligned}
    &0.0496^{+0.0019}_{-0.0071}~ \Lambda_\text{F},\text{ fermionic;} \\
    &0.0444^{+0.0033}_{-0.0015}~ \Lambda_\text{B},\text{ bosonic.}
    \end{aligned}\right.
\end{equation}
Here, $\Lambda_\text{F}$ ($\Lambda_\text{B}$) is the combined dimensionless \ac{DM} parameter for self-annihilating \ac{FDM} (\ac{BDM}) given by
\begin{align}
\Lambda_\text{F} &=
\left(\frac{\braket{\sigma v}}{10^{-42}~\si{cm}^3~\si{s}^{-1}}\right)\left(\frac{m_\chi}{0.1~\si{GeV}}\right)\left(\frac{f_d}{10^{-8}}\right)^{1.4},\\
\Lambda_\text{B} &= \Lambda_\text{F} \left(\frac{1~\si{pm}}{a_s}\right)^{0.6},
\end{align}
where $f_d=M_d/M_t$. $a_s$ is typically about $1~\si{pm}$ for the \ac{BDM} to provide the same order of heating effect as \ac{FDM}.

Details of the fitting are presented in TABLE~\ref{tab:L fitting range} in Appendix~\ref{app:fitting}, with fitted results presented in TABLE~\ref{tab:L fitting stats}, showing a reasonably good approximation to the \ac{SADM}-admixed \ac{WD} luminosity arising from \ac{SADM} (see FIG.~\ref{fig:L fitting}).

The total luminosity could be expressed analytically as
\begin{equation}
    L=\braket{\sigma v}f_dM_t\frac{\braket{n_d^2}_V}{\braket{n_d}_V},
\end{equation}
where $n_d$ is the number density of \ac{DM} particles, and the angled bracket $\braket{\cdot}_V$ denotes the volume-average.

From our numerical solution of the hydrostatic equations, we found that, to a good approximation,
\begin{equation}
    \Lambda M_t^{3.5}\propto \braket{\sigma v}f_dM_t\frac{\braket{n_d^2}_V}{\braket{n_d}_V},
\end{equation}
which implies $L\propto \Lambda M_t^{3.5}$, for both $\Lambda=\Lambda_\text{F}$ and $\Lambda_\text{B}$. Therefore, the reported fitting formula resembles our understanding of the luminosity.

 
\subsection{Empirical fitting of $T_\text{eff}$}

We also explore the relationship between the effective surface temperature $T_\text{eff}$ with $L$ and $\Lambda_\text{F}$ ($\Lambda_\text{B}$) for self-annihilating \ac{FDM} (\ac{BDM}) admixed \acp{WD}. We obtain the empirical fitting formula
\begin{equation} \label{eq:Teff fitting}
    T_\text{eff}= \alpha~\left(\frac{L}{0.1~L_\odot~\Lambda^{1/3}}\right)^{0.36},
\end{equation}
where $\alpha=38977^{+690}_{-377}~\si{K}~\left(38333^{+561}_{-451}~\si{K}\right)$ and $\Lambda=\Lambda_\text{F}~(\Lambda_\text{B})$ for \ac{FDM} (\ac{BDM}).

We obtain Eq.~\eqref{eq:Teff fitting} for the same range of parameters listed in TABLE~\ref{tab:L fitting range}. The fittings for both self-annihilating \ac{FDM} and \ac{BDM} are shown as the red dashed lines in the upper and lower panels in FIG.~\ref{fig:Teff fitting}, respectively, with the fitting statistics reported in TABLE~\ref{tab:Teff fitting stats}. The empirical fitting, Eq.~\eqref{eq:Teff fitting}, provides a reasonably similar estimate of the effective surface temperature $T_\text{eff}$ for both the self-annihilating \ac{FDM} and \ac{BDM} admixed \acp{WD}.


\subsection{\ac{MR} relation for \ac{SADM}-admixed \ac{WD}}

\begin{figure}
\includegraphics[width=\linewidth]{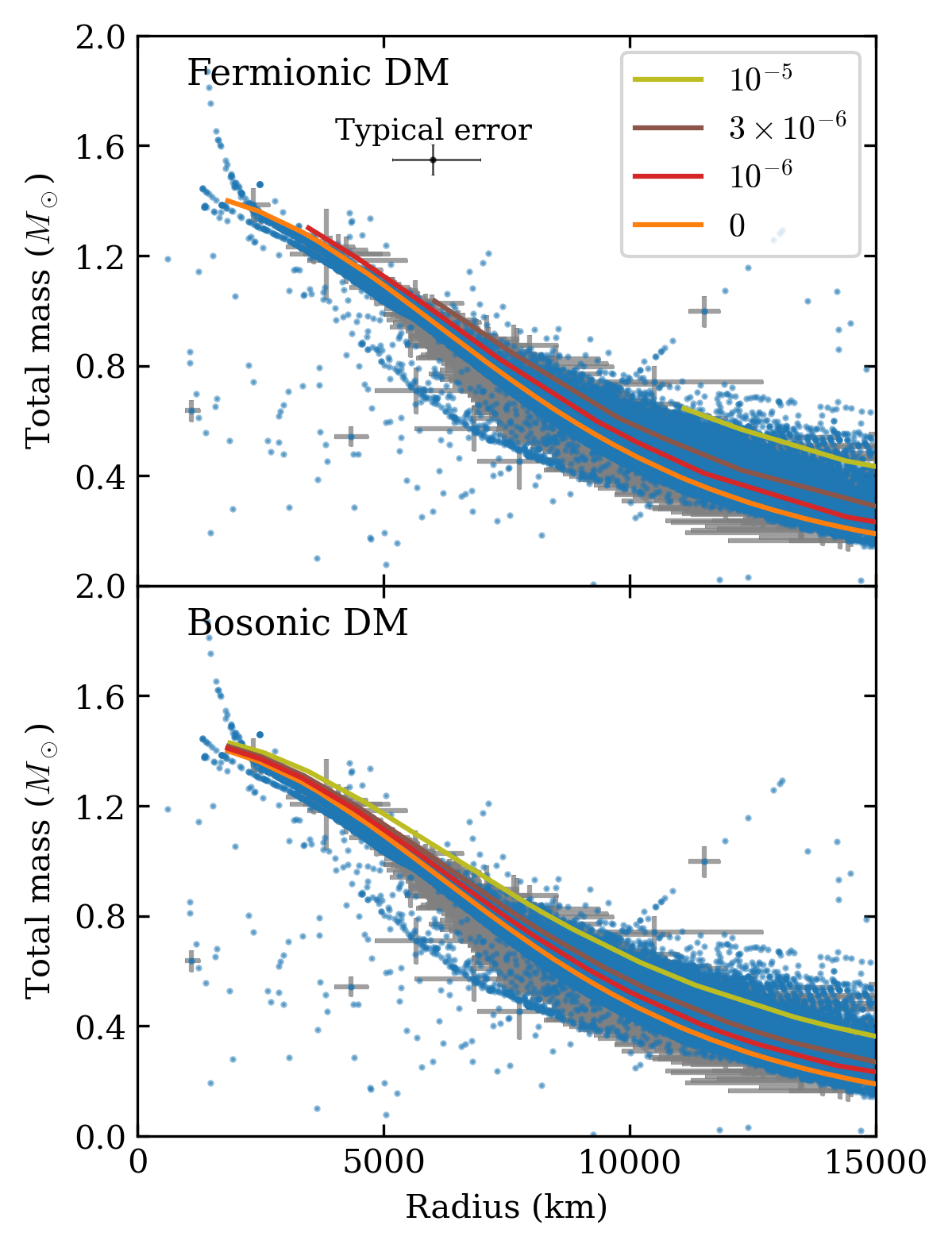}
\caption{\label{fig:f_d MR} Same as FIG.~\ref{fig:MRT}, but the observational data are in blue, together with typical error bars for a small subset of data. The upper (lower) panel is for self-annihilating \ac{FDM} (\ac{BDM}). Those with \ac{DM} mass fraction $f_d=10^{-5}$, $3\times 10^{-6}$, $10^{-6}$, and $0$ are plotted in olive, brown, red, and orange, respectively. The cross section $\braket{\sigma v}$ is $10^{-44}$ cm$^3$ s$^{-1}$ and the DM particle mass $m_\chi$ is 0.1 GeV. The scattering length $a_s$ for the \ac{BDM} in the lower panel is $1$ pm.}
\end{figure}

We plot the \ac{SADM}-admixed \ac{WD} \ac{MR} relations for different $f_d$ for \ac{FDM} (\ac{BDM}) in the upper (lower) panel, in FIG.~\ref{fig:f_d MR}, to show that a tiny mass fraction of \ac{SADM} ($f_d\in[10^{-6},~10^{-5}]$), either \ac{FDM} or \ac{BDM}, with a small $\braket{\sigma v}$, can account for the majority of \acp{WD} with \ac{MR} relation deviating from the theoretical \ac{MR} relation with no \ac{DM} admixed (shown as the orange line).

\begin{figure}
\includegraphics[width=\linewidth]{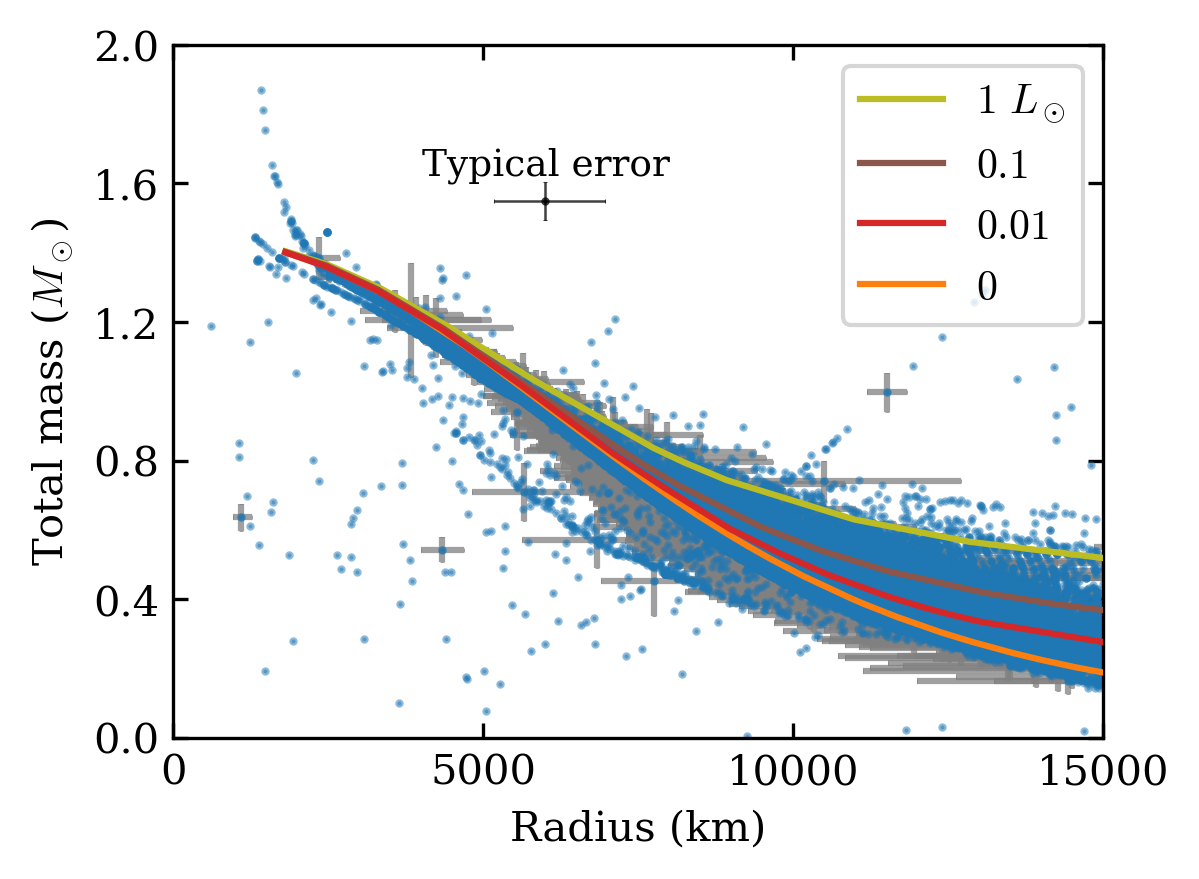}
\caption{\label{fig:L MR}Same as FIG.~\ref{fig:f_d MR}, but for \ac{SADM}-admixed \acp{WD} with luminosity $L=0.01,~0.1,~1~L_\odot$ generated from \ac{SADM}, plotted in red, brown, and olive, respectively. Both \ac{FDM} and \ac{BDM} show similar shifts in the \ac{MR} relation.}
\end{figure}

We further plot the \ac{MR} relations for \ac{SADM}-admixed \acp{WD} with different $L$ in FIG.~\ref{fig:L MR}. Both \ac{FDM} and \ac{BDM} show similar shifts in the \ac{MR} relation. With a larger $L$, the \ac{WD} temperature is higher, resulting in a larger thermal pressure to support the star. This explains why the \ac{MR} relation shifts upward with increasing luminosity. 

Note that for the same $L$, the shift in the \ac{MR} relation is more significant for lower-mass \acp{WD}, which are less compact, so the thermal pressure accounts for a larger fractional increase in the supportive force against gravity.


\section{Discussion and Conclusion} \label{sec:Conclusion}

We have presented a new framework for a \ac{SADM}-admixed \acp{WD} using a hydrostatic two-fluid model, with the \ac{DM} as a core comprising degenerate \ac{FDM} or Bose-Einstein condensate of \ac{BDM}, embedded in the \acp{WD}. The \ac{DM} self-annihilates into photons, which diffuse and heat the \ac{WD}. We have modeled the \ac{SADM}-admixed \acp{WD} with broad parameter ranges, reported in TABLE~\ref{tab:L fitting range}. With the new framework, we can explore the impact of the energy generation from \ac{SADM} on the \ac{WD} structure.

Our study shows that, typically, for the \ac{SADM} with $m_\chi=0.1~\si{GeV}$, the \ac{DM} annihilation cross section could be about $10^{-44}~\si{cm}^3~\si{s}^{-1}$, which is $18$ orders of magnitude lower than the thermal relics cross section \cite{PhysRevD.86.023506}. In addition, the observables of \ac{SADM}-admixed \acp{WD} depend on one combination of \ac{DM} parameters, $\Lambda_\text{F}$ or $\Lambda_\text{B}$. We show that \acp{WD} could be a sensitive laboratory to probe the quantity $\braket{\sigma v}m_\chi$ for \ac{FDM} ($\frac{\braket{\sigma v}m_\chi}{a_s}$ for \ac{BDM}).

Furthermore, the \ac{SADM}-admixed \acp{WD}' luminosity $L$ and effective temperature $T_\text{eff}$ could be represented by empirical formulae. For both fermionic and bosonic \ac{SADM}-admixed \acp{WD}, there is a systematic correlation between $L$ and the shift of the \ac{MR} relation. Moreover, assuming benchmark values of $\braket{\sigma v}=10^{-44}~\si{cm}^3~\si{s}^{-1}$ and $m_\chi=0.1~\si{GeV}$, a small fraction ($f_d\in[10^{-6},~10^{-5}]$) of \ac{SADM} could account for the unknown energy source, leading to an anomalously high temperature, luminosity, and deviation from the theoretical no-\ac{DM} admixed \ac{MR} relation observed for many \acp{WD}.

While this work focuses on spherically symmetric configurations, stellar rotation is expected to deform the structure \cite{1968ApJ...153..807H, 1985A&A...146..260E, 1986ApJS...62..461H, Chan_2022} and alter the heating profile of \ac{SADM}-admixed \acp{WD}. Rotational effects will be presented in a forthcoming paper.

Our framework could be extended further by considering other energy production processes inside the \acp{WD}. One can apply this for decaying \ac{DM}, primordial black holes \cite{AUFFINGER2023104040, PhysRevD.92.063007}, dark photons \cite{fabbrichesi2021physics}, and other similar \ac{DM} candidates, by using the corresponding energy production rate. This could also be extended to neutron stars, and the admixed \ac{SADM} may serve as a portal to constrain the nuclear matter \ac{EoS}.

It is also possible for \ac{SADM}-admixed \acp{WD} to reach a high enough temperature from \ac{DM} self-annihilation to trigger Carbon fusion, thus exploding the star \cite{PhysRevD.98.115027, PhysRevLett.115.141301}. This suggests a new type of transient events, the detail modeling of which is an interesting future work.

\acp{WD} are abundant in the Milky Way and other galaxies, and there will be many more high-precision observations of \acp{WD} in the near future. This work suggests that \acp{WD} can serve as sensitive detectors for self-annihilating dark matter.


\begin{acknowledgments}
This research project is partially supported by grants from the Research Grant Council of the Hong Kong Special Administrative Region, China (Project Nos. 14300320 and 14304322). P. S-H. C. acknowledges funding by the European Union (ERC, ExCEED, project number 101096243). Views and opinions expressed are however those of the authors only and do not necessarily reflect those of the European Union or the European Research Council Executive Agency. Neither the European Union nor the granting authority can be held responsible for them. We acknowledge support by the Klaus Tschira Foundation.
\end{acknowledgments}


\begin{appendix}

\section{Formation of \ac{DM} admixture} \label{app:capture rate}
\ac{SADM} could reside in the \ac{MS} stars, prior to the formation of \acp{WD} \cite{moskalenko2007darkmatterburnerspreliminary, Iocco_2008}. In this estimation, we consider a typical $1~M_\odot$ \ac{WD}, together with the corresponding $5~M_\odot$ \ac{MS} phase \cite{CummingIFMR}, with a radius of $2.5~R_\odot$ \cite{Demircan1991}, and a $0.1~\si{Gyr}$ lifetime \cite{MIST}. For a $0.1~\si{GeV}$ \ac{DM} particle with the \ac{DM}-nucleon interaction cross section as $10^{-31}~\si{cm}^2$ \cite{PhysRevD.104.012009, PhysRevLett.122.171801, PhysRevD.100.103011}, both the \ac{WD} and the \ac{MS} phases are optically thick, so we compute the capture rate $C$ using Eq.~(8) in \cite{moskalenko2007darkmatterburnerspreliminary} 
\begin{equation}
C=\left(\frac{3}{8\pi}\right)^{1/2} \frac{\rho_\chi \bar{v}}{m_\chi}\left(\zeta+\frac{3 v_\text{esc}^2}{2\bar{v}^2}\right) \pi R_n^2,
\end{equation}
where $\rho_\chi$ is the surrounding \ac{DM} density, $\bar{v}$ is the \ac{DM} velocity dispersion, $\zeta=1.77$, and $v_\text{esc}=\sqrt{2GM_t/R_n}$ is the escape velocity of the star.

In the solar neighbourhood ($\rho_\chi=0.43~\si{GeV}~\si{cm}^{-3}$, $\bar{v}=215~\si{km}~\si{s}^{-1}$) \cite{Solarneighbourhood}, the \ac{DM} is captured at the rate of about $10^{-12}~M_\odot~\si{Gyr}^{-1}$ ($10^{-7}~M_\odot~\si{Gyr}^{-1}$) for the $1~M_\odot$ \ac{WD} ($5~M_\odot$ \ac{MS} star). Therefore, the \ac{MS} phase dominates the amount of \ac{DM} captured. We further assume the Navarro-Frenk-White (NFW) \ac{DM} density profile \cite{NFW} with standard Milky Way halo parameters \cite{10.1093/mnras/stw2759}, and at about $0.1~\si{pc}$ from the galactic centre ($\rho_\chi=10^5~\si{GeV}~\si{cm}^{-3}$, $\bar{v}=300~\si{km}~\si{s}^{-1}$ \cite{2009A&A...502...91S}), the \ac{MS} star can capture \ac{DM} at a rate of $10^{-4}~M_\odot~\si{Gyr}^{-1}$.

Therefore, the \ac{MS} star can capture a total of $10^{-5}~M_\odot$ \ac{DM} within $0.1~\si{Gyr}$. The actual captured \ac{DM} mass can be further boosted with a denser \ac{DM} surrounding and a larger \ac{MS} progenitor. The parameter range we considered in this study is therefore physically possible.


\section{Luminosity and effective temperature fitting} \label{app:fitting}
The fitted parameter ranges are reported in TABLE~\ref{tab:L fitting range}. The statistics for luminosity and effective temperature fittings are listed in TABLE~\ref{tab:L fitting stats} and TABLE~\ref{tab:Teff fitting stats}, respectively.

\begin{table}
    \centering
    \renewcommand{\arraystretch}{1.5}
    \hspace*{-0.5cm}
    \begin{tabular}{c|c|c}
    \hline\hline
        & Fermionic & Bosonic\\
    \hline
        $L~(L_\odot)$ & $[4\times10^{-7},~10]$ & $[10^{-6},~4]$\\
        \hline
        $f_d$ & $[4\times10^{-15},~4\times10^{-3}]$ & $[10^{-17},~8\times10^{-3}]$  \\
        \hline
        $\braket{\sigma v}~(\si{cm}^3~\si{s}^{-1})$ & $[10^{-48},~10^{-36}]$ & $[10^{-47},~10^{-34}]$ \\
        \hline
        $m_\chi~(\si{GeV})$& $[0.01,~10]$& $[0.01,~7]$\\
        \hline
        $M_t~(M_\odot)$& $[0.150,~1.403]$ & $[0.058,~1.366]$\\
        \hline
        $a_s~(\si{pm})$& /&$[0.0002,~4000]$\\
    \hline\hline
    \end{tabular}
    \caption{\label{tab:L fitting range}Parameter range for empirical fitting of \ac{WD} luminosity $L$ and effective temperature $T_\text{eff}$ from self-annihilating \ac{DM}.}
\end{table}

\begin{table}
    \centering
    \renewcommand{\arraystretch}{1.5}
    \begin{tabular}{c|c|c}
    \hline\hline
        & Fermionic & Bosonic\\
    \hline
        Prefactor & $0.0496^{+0.0019}_{-0.0071}$ & $0.0444^{+0.0033}_{-0.0015}$\\
        \hline
        Reduced chi-squared $\chi^2_\nu$ & $0.9891$ & $1.0077$  \\
        \hline
        Bin count & $19$ & $105$ \\
        \hline
        Degree of freedom & $18$& $104$\\
    \hline\hline
    \end{tabular}
    \caption{\label{tab:L fitting stats}Statistics of the empirical fitting of luminosity $L$ from self-annihilating \ac{DM}.}    
\end{table}

\begin{figure}
    \centering
    \includegraphics[width=\linewidth]{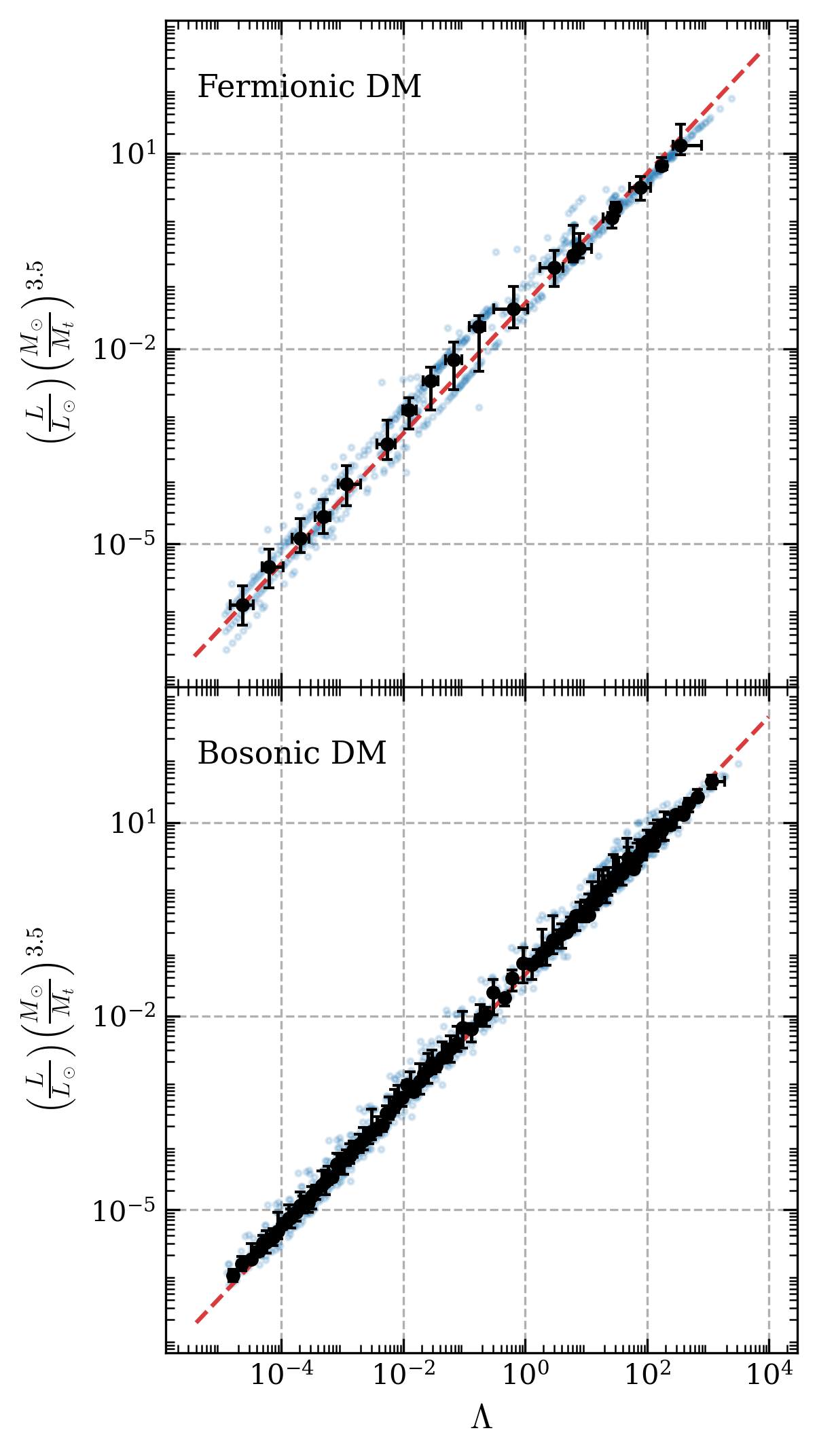}
    \caption{\label{fig:L fitting}Total luminosity of \ac{SADM}-admixed \acp{WD} in solar luminosity fitting with \ac{WD} mass $M_t$ and \ac{DM} parameter $\Lambda=\Lambda_\text{F}$ ($\Lambda_\text{B}$) for self-annihilating \ac{FDM} (\ac{BDM}) in the upper (lower) panel. The faded blue dots are numerical results, and the black dots are the median values of each bin data bin, while the error bars report the $16^\text{th}$ and $84^\text{th}$ percentiles. The binned data are well represented by our fitting formula (dashed line).}
\end{figure}

The fittings of \ac{SADM}-admixed \ac{WD} luminosity $L$ and total mass $M_t$ against the \ac{DM} parameters are shown in FIG.~\ref{fig:L fitting}.

\begin{table}[h]
    \centering
    \renewcommand{\arraystretch}{1.5}
    \begin{tabular}{c|c|c}
    \hline\hline
        & Fermionic & Bosonic\\
    \hline
        Prefactor & $38977^{+690}_{-377}$ & $38333^{+561}_{-451}$\\
        \hline
        Reduced chi-squared $\chi^2_\nu$ & $1.0947$ & $1.5674$  \\
        \hline
        Bin count & $34$ & $51$ \\
        \hline
        Degree of freedom & $33$& $50$\\
    \hline\hline
    \end{tabular}
    \caption{\label{tab:Teff fitting stats}Same as TABLE~\ref{tab:L fitting stats}, but for $T_\text{eff}$.}
\end{table}

\begin{figure}
    \centering
    \includegraphics[width=\linewidth]{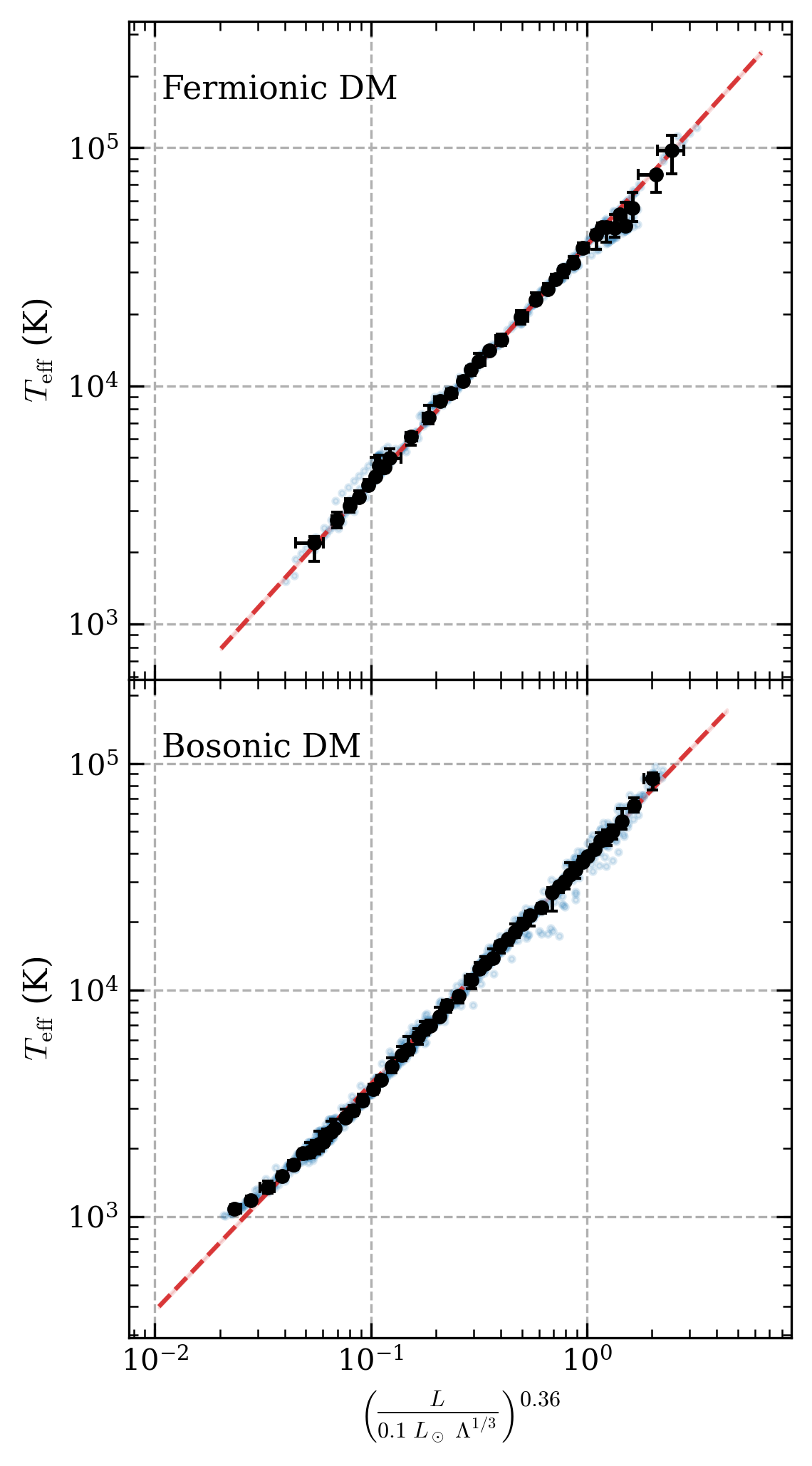}
    \caption{\label{fig:Teff fitting}Same as FIG.~\ref{fig:L fitting}, but for $T_\text{eff}$ vs. $\left(\frac{L}{0.1~L_\odot~\Lambda^{1/3}}\right)^{0.36}$.}
\end{figure}

The fittings of \ac{SADM}-admixed \ac{WD} effective temperature $T_\text{eff}$ against the luminosity $L$ and the \ac{DM} parameters are shown in FIG.~\ref{fig:Teff fitting}.

\end{appendix}

\bibliography{SADM}

\end{document}